\documentclass[twocolumn]{aastex631}
\usepackage{amsfonts}
\usepackage{mathrsfs}
\usepackage{natbib}
\usepackage{hyperref}
\usepackage{bm}
\usepackage{amsmath}
\DeclareMathAlphabet{\mathscrbf}{OMS}{mdugm}{b}{n}
\usepackage{color}
\usepackage{acro}
\usepackage{svg}
\acsetup{patch/longtable=false}

\DeclareAcronym{GW}{
  short = GW ,
  long = gravitational wave ,
  short-plural = s 
}
\DeclareAcronym{LIGO}{
  short = LIGO ,
  long = Laser Interferometer Gravitational-wave Observatory ,
  short-plural = 
}
\DeclareAcronym{LISA}{
  short = LISA ,
  long = Laser Interferometer Space Antenna ,
  short-plural =  
}
\DeclareAcronym{SKA}{
  short = SKA ,
  long = Square Kilometre Array ,
  short-plural =  
}  
  
\DeclareAcronym{SNR}{
	short = SNR ,
	long = signal-to-noise ratio ,
	short-plural = 
}

\DeclareAcronym{PTA}{
	short = PTA ,
	long = pulsar timing array ,
	short-plural = 
}

\DeclareAcronym{FLRW}{
  short = FLRW ,
  long = Friedmann-Lemaitre-Robertson-Walker ,
  short-plural =  
}

\DeclareAcronym{SIGW}{
	short = SIGW ,
	long = scalar induced gravitational wave ,
	short-plural =  s
}

\DeclareAcronym{PBH}{
	short = PBH ,
	long = primordial black hole ,
	short-plural =  s
}

\DeclareAcronym{CMB}{
	short = CMB ,
	long = cosmic microwave background ,
	short-plural =  
}
\DeclareAcronym{DM}{
	short = DM ,
	long = dark matter ,
	short-plural =  
}

\DeclareAcronym{SGWB}{
	short = SGWB ,
	long = stochastic gravitational	wave background ,
	short-plural =  s
}

\DeclareAcronym{LSS}{
	short = LSS ,
	long = large scale structure ,
	short-plural =  
}

\DeclareAcronym{RD}{
	short = RD ,
	long = radiation-dominated ,
	short-plural =  
}

\DeclareAcronym{BAO}{
	short = BAO ,
	long = baryon acoustic oscillations ,
	short-plural = 
}

\begin{document}

\title{Constraints on primordial black holes from $N_{\text{eff}}$ : scalar induced gravitational waves as an extra radiation component}

\author{Jing-Zhi Zhou} 
\email{zhoujingzhi@tju.edu.cn}
\affiliation{Center for Joint Quantum Studies and Department of Physics,
School of Science, Tianjin University, Tianjin 300350, China}

\author{Yu-Ting Kuang} 
\affiliation{Institute of High Energy Physics, Chinese Academy of Sciences, Beijing 100049, China}
\affiliation{University of Chinese Academy of Sciences, Beijing 100049, China}

\author{Zhe Chang} 
\affiliation{Institute of High Energy Physics, Chinese Academy of Sciences, Beijing 100049, China}
\affiliation{University of Chinese Academy of Sciences, Beijing 100049, China}

\author{ H. L\"u}
\affiliation{Center for Joint Quantum Studies and Department of Physics,
School of Science, Tianjin University, Tianjin 300350, China}
\affiliation{Joint School of National University of Singapore and Tianjin University, International Campus of Tianjin University, Binhai New City, Fuzhou 350207, China}

\begin{abstract}
In June 2023, multiple pulsar timing array collaborations provided evidence for the existence of a stochastic gravitational wave background. Scalar induced gravitational waves (SIGWs), as one of the most likely sources of stochastic gravitational waves, have received widespread attention. When primordial curvature perturbations on small scales are sufficiently large, \acp{PBH} inevitably form, concurrently producing SIGWs with significant observable effects. These SIGWs can serve as an additional radiation component, influencing the relativistic degrees of freedom $N_{\text{eff}}$.  Taking into account primordial non-Gaussianity, we study the energy density spectrum of SIGWs up to the third order and use the current observational data of $N_{\text{eff}}$ to constrain small-scale primordial curvature perturbations and the abundance of \acp{PBH}. 
\end{abstract}

\keywords{Primordial black hole --- Scalar induced gravitational wave --- Primordial curvature perturbation --- Primordial non-Gaussianity}

\section{Introduction} \label{sec:1.0}
Current cosmological observations suggest the existence of dark matter in our universe. As one of the significant candidates for dark matter, \acp{PBH} have garnered extensive attention over the past decade. \acp{PBH} originate from primordial curvature perturbations with large amplitudes on small scales \cite{Bird:2016dcv,Garcia-Bellido:2017fdg,Barack:2018yly,Byrnes:2018txb,Chen:2024pge}. Specifically, on large scales ($\gtrsim$1 Mpc), cosmological observations like the \acp{CMB} and \acp{LSS} have tightly constrained the amplitude of the primordial curvature perturbation power spectrum to $A_{\zeta}\sim 2\times 10^{-9}$ \cite{Planck:2018vyg,Abdalla:2022yfr}. The tensor-to-scalar ratio $r$ on large scales is restricted to below $0.06$. However, on small scales ($\lesssim$1 Mpc), there are no particularly stringent experimental observational constraints on primordial curvature perturbations \cite{Bringmann:2011ut}. This enables large amplitude primordial perturbations on small scales to re-enter the horizon after inflation, resulting in significant density perturbations and the formation of \acp{PBH} \cite{Sasaki:2018dmp,Carr:2020gox}. 

The process during which the \acp{PBH} are formed would be inevitably accompanied by the generation of SIGWs \cite{Matarrese:1997ay,Mollerach:2003nq,Ananda:2006af,Baumann:2007zm,Saito:2008jc,Bugaev:2009zh,Bugaev:2010bb,Yuan:2019udt,Papanikolaou:2020qtd,Chang:2022nzu,Choudhury:2024dzw,Domenech:2024wao}. In June 2023, multiple international PTA collaborations, including NANOGrav \cite{NANOGrav:2023gor}, PPTA \cite{Reardon:2023gzh}, EPTA \cite{EPTA:2023fyk}, and CPTA \cite{Xu:2023wog}, presented evidence supporting the existence of a \acp{SGWB} in the nHz frequency range. SIGWs, as one of the most likely sources of the stochastic gravitational wave background, have garnered significant attention \cite{Franciolini:2023pbf,Ellis:2023oxs,Domenech:2024rks,HosseiniMansoori:2023mqh,Harigaya:2023pmw}. Utilizing current \ac{PTA} data, we can investigate the impact of SIGWs on current \acp{SGWB}, thereby constraining the parameter space of the primordial power spectrum on small scales and the abundance of \acp{PBH}. Furthermore, current \ac{PTA} data can be used to investigate potential new physics that might have existed during the evolution of the universe \cite{Yu:2024xmz,Sui:2024nip,Zhou:2024doz,Papanikolaou:2024fzf,Papanikolaou:2023oxq}.

In this paper, we consider the impact of SIGWs with primordial non-Gaussianity as an additional radiation component on cosmological evolution. More precisely, current large-scale cosmological observations, such as \ac{CMB}, \ac{LSS}, and \ac{BAO} provide stringent constraints on the relativistic degrees of freedom parameter $N_{\text{eff}}$ \cite{Jiang:2023gfe,Ben-Dayan:2019gll,Clarke:2020bil,Cang:2022jyc,Wang:2023sij}. When we consider the additional contribution of SIGWs to the radiation components, this additional radiation component will inevitably affect the parameter $N_{\text{eff}}$, and this effect must not exceed the bounds on  $N_{\text{eff}}$ determined by current large-scale cosmological observations.  Utilizing the current observational data of $N_{\text{eff}}$, we can constrain the energy density spectrum of SIGWs, thereby limiting and excluding the parameter space of the primordial power spectrum on small scales. Since the primordial power spectrum directly influences the calculation of the abundance of \acp{PBH}, this physical process can also be used to constrain the abundance of \acp{PBH} of different masses.

 This paper is organized as follows. In Sec.~\ref{sec:2.0}, we review the calculations of second and third order SIGWs. In Sec.~\ref{sec:3.0}, we calculate the energy density spectrum of SIGWs up to third order. In Sec.~\ref{sec:4.0}, we discuss the constraints on the abundance of \acp{PBH} from large-scale cosmological observations. Finally, we summarize our results and give some discussions in Sec.~\ref{sec:5.0}.

\section{Second and third order SIGWs } \label{sec:2.0}
Due to the potential presence of significant primordial curvature perturbations on small scales, higher-order SIGWs can have a substantial impact on the total energy density spectrum of SIGWs. Ignoring the higher-order contributions of SIGWs will lead to severe deviations in the predicted energy density spectrum of SIGWs. Therefore, it is crucial to consider the impact of higher-order effects on the energy density spectrum of SIGWs. In this section, we briefly review the main results of second order and third order SIGWs. The perturbed metric in the \ac{FLRW} spacetime with
Newtonian gauge takes the form
\begin{equation}\label{eq:ds}
	\begin{aligned}
		&\mathrm{d}s^{2}=a^{2}\left(-\left(1+2 \phi^{(1)}+ \phi^{(2)}\right) \mathrm{d} \eta^{2}+ V_i^{(2)} \mathrm{d} \eta \mathrm{d} x^{i}+\right. \\
		&\left.\left(\left(1-2 \psi^{(1)}- \psi^{(2)}\right) \delta_{i j}+\frac{1}{2} h_{i j}^{(2)}+\frac{1}{6} h_{i j}^{(3)}\right)\mathrm{d} x^{i} \mathrm{d} x^{j}\right) \ ,
	\end{aligned}
\end{equation}
where $h^{(n)}_{ij}$$\left( n=2,3 \right)$ are second order and third order tensor perturbations. $\phi^{(n)}$ and $\psi^{(n)}$$\left( n=1,2 \right)$ are first order and second order scalar perturbations. $V_i^{(2)} $ is second order vector perturbation. By substituting the metric perturbations in Eq.~(\ref{eq:ds}) into the Einstein field equation and simplifying, we can derive the equations of motion of the cosmological perturbations at each order \cite{Zhou:2024ncc}. To obtain the explicit expressions for the second order and third order SIGWs, we need to start with the first order scalar perturbations and solve these cosmological perturbation equations order by order. After solving the first order scalar perturbations, we can derive the specific expression for the second order SIGWs \cite{Kohri:2018awv}
\begin{equation}
	\begin{aligned}
		h^{\lambda,(2)}_{\mathbf{k}}(\eta)= \frac{4}{9}\int \frac{d^3 p}{(2 \pi)^{3 / 2}}  \varepsilon^{\lambda, l m}(\mathbf{k})p_l p_m  \zeta_{\mathbf{k}-\mathbf{p}}  \zeta_{\mathbf{p}} I^{(2)}(u,v) \ ,
	\end{aligned}\label{eq:2h}
\end{equation}
where 
\begin{equation}
\begin{aligned}
I^{(2)}(u, v)&=\frac{27}{64}\left(\frac{3\left(u^2+v^2-3\right)\left(-4 v^2+\left(1-u^2+v^2\right)^2\right)}{16 u^4 v^4}\right)^2 \\
&\left(\left(-4 u v+\left(u^2+v^2-3\right) \log \left|\frac{3-(u+v)^2}{3-(u-v)^2}\right|\right)^2\right. \\
&\left.+\pi^2\left(u^2+v^2-3\right)^2 \Theta(u+v-\sqrt{3})\right)
\end{aligned}
\end{equation}
is known as the kernel function of second order SIGWs. Here, we have defined $|\mathbf{k}-\mathbf{p}|=u|\mathbf{k}|$ and $|\mathbf{p}|=v|\mathbf{k}|$.

For third order SIGWs, first order scalar perturbations directly induce three types of second order perturbations. These three second order perturbations, together with the first order scalar perturbations, subsequently induce third order SIGWs \cite{Chang:2023vjk,Zhou:2021vcw}. The specific expression for the third order SIGWs can be represented as
\begin{eqnarray}
	h^{\lambda,(3)}_{\mathbf{k}}(\eta)& &=h^{\lambda,(3)}_{\mathbf{k},\phi\phi\phi}(\eta)+h^{\lambda,(3)}_{\mathbf{k},\phi h_{\phi\phi}}(\eta)+h^{\lambda,(3)}_{\mathbf{k},\phi V_{\phi\phi}}(\eta) \nonumber\\
	& &+h^{\lambda,(3)}_{\mathbf{k},\phi \psi_{\phi\phi}}(\eta) \  ,
	\label{eq:H0}
\end{eqnarray}
where
\begin{align}
	h^{\lambda,(3)}_{\mathbf{k},\phi\phi\phi}&(\eta)=\int\frac{d^3p}{(2\pi)^{3/2}}\int\frac{d^3q}{(2\pi)^{3/2}}\varepsilon^{\lambda,lm}(\mathbf{k})q_m \nonumber\\
	&\times (p_l-q_l)\frac{8}{27} I_{\phi\phi\phi}^{(3)}(u,\bar{u},\bar{v},x)\zeta_{\mathbf{k}-\mathbf{p}} \zeta_{\mathbf{p}-\mathbf{q}} \zeta_{\mathbf{q}} \  ,
	\label{eq:H1}
 \end{align}
 \begin{align}
	h^{\lambda,(3)}_{\mathbf{k},\phi h_{\phi\phi}}&(\eta)=\int\frac{d^3p}{(2\pi)^{3/2}}\int\frac{d^3q}{(2\pi)^{3/2}} \varepsilon^{\lambda, lm}(\mathbf{k})\Lambda_{lm}^{ rs}(\mathbf{p})\nonumber\\
	&\times
	q_rq_s\frac{8}{27}I^{(3)}_{\phi h_{\phi\phi}}(u,\bar{u},\bar{v},x)\zeta_{\mathbf{k}-\mathbf{p}} \zeta_{\mathbf{p}-\mathbf{q}} \zeta_{\mathbf{q}} \ ,
	\label{eq:H2} 
  \end{align}
 \begin{align}
	h^{\lambda,(3)}_{\mathbf{k},\phi V_{\phi\phi}}&(\eta)=\int\frac{d^3p}{(2\pi)^{3/2}}\int\frac{d^3q}{(2\pi)^{3/2}}\varepsilon^{\lambda,lm}(\mathbf{k})\mathcal{T}^r_{(m}(\textbf{p}) p_{l)}  \nonumber\\
	&\times  \frac{16p^s}{27p^2}q_rq_sI^{(3)}_{\phi V_{\phi\phi}}(u,\bar{u},\bar{v},x)\zeta_{\mathbf{k}-\mathbf{p}}\zeta_{\mathbf{p}-\mathbf{q}} \zeta_{\mathbf{q}} \ ,
	\label{eq:H3}
  \end{align}
 \begin{align}
	h^{\lambda,(3)}_{\mathbf{k},\phi \psi_{\phi\phi}}&(\eta)=\int\frac{d^3p}{(2\pi)^{3/2}}\int\frac{d^3q}{(2\pi)^{3/2}}\varepsilon^{\lambda,lm}(\mathbf{k})p_lp_m \nonumber\\
	&\times
	\frac{8}{27}I^{(3)}_{\phi \psi_{\phi\phi}}(u,\bar{u},\bar{v},x)\zeta_{\mathbf{k}-\mathbf{p}} \zeta_{\mathbf{p}-\mathbf{q}} \zeta_{\mathbf{q}} \ .
	\label{eq:H4}
\end{align}
Here, we have defined $x=|\mathbf{k}|\eta$. In Eq.~(\ref{eq:H1}), we use the symbol $h^{(3)}_{\mathbf{k},\phi\phi\phi}$ to represent the third order SIGWs sourced by three first order scalar perturbations \cite{Chen:2022dah}. Eq.~(\ref{eq:H2})--Eq.~(\ref{eq:H4}) represent third order SIGWs induced by three types of second order perturbations together with first order scalar perturbations. The explicit expressions of the third order kernel functions $I^{(3)}$ can be found in Ref.~\cite{Zhou:2021vcw}.

\section{Energy density of SIGWs} \label{sec:3.0}
In this section, we utilize the specific expressions for second order and third order SIGWs provided in the previous section to study the energy density spectra of SIGWs. The energy density spectrum of SIGWs is defined as
\begin{equation}\label{eq:Ome}
    \overline{\Omega}_{\mathrm{GW}}(\eta, k) = \overline{\frac{\rho_{\mathrm{GW}}(\eta,k)}{\rho_{\mathrm{tot}}(\eta)}} = \overline{\frac{x^2}{6}\mathcal{P}_h(\eta,k)}  \ ,
\end{equation}
where
\begin{eqnarray}
	 \mathcal{P}_h(\eta,k) &=& \frac{1}{4}\mathcal{P}^{(2,2)}_h(\eta,k) + \frac{1}{6}\mathcal{P}^{(2,3)}_h(\eta,k)+ \frac{1}{36}\mathcal{P}^{(3,3)}_h(\eta,k) \nonumber\\
    &+&O(\mathcal{P}^{(4,n)}_h) 
\end{eqnarray}
is the total power spectrum of SIGWs. The overline represents the oscillation average. The power spectrum $\mathcal{P}^{(m,n)}_k(\eta)$ can be calculated in terms of the two-point correlation function of $h^{\lambda,(m)}_{\mathbf{k}}(\eta)$ and $h^{\lambda,(n)}_{\mathbf{k}}(\eta)$
\begin{equation}\label{eq:Ph} 
	\begin{aligned}
		\langle h^{\lambda,(m)}_{\mathbf{k}}(\eta) h^{\lambda^{\prime},(n)}_{\mathbf{k}'}(\eta)\rangle= \delta^{\lambda \lambda'}\delta\left(\mathbf{k}+\mathbf{k}'\right) \frac{2 \pi^{2}}{k^{3}} \mathcal{P}^{(m,n)}_k(\eta) \ .
	\end{aligned}
\end{equation}
Before calculating the total energy density spectrum of SIGWs, we can first analyze the structure of the two-point correlation function $\langle h^{\lambda,(m)}_{\mathbf{k}}(\eta) h^{\lambda^{\prime},(n)}_{\mathbf{k}'}(\eta)\rangle$. Specifically, as shown in Eq.~(\ref{eq:H0})--Eq.~(\ref{eq:H4}) and Eq.~(\ref{eq:2h}), the second order SIGW is proportional to two primordial curvature perturbations: $h^{\lambda,(2)}_{\mathbf{k}}(\eta)\sim\zeta_{\mathbf{k}-\mathbf{p}}\zeta_{\mathbf{p}}$, and the third order SIGW is proportional to three primordial curvature perturbations: $h^{\lambda,(3)}_{\mathbf{k}}(\eta)\sim\zeta_{\mathbf{k}-\mathbf{p}}\zeta_{\mathbf{p}-\mathbf{q}}\zeta_{\mathbf{q}}$. When we only consider Gaussian-type primordial curvature perturbations, the $n$-point correlation function of the primordial curvature perturbations is strictly zero when $n$ is odd. Therefore, the two-point correlation function $\langle h^{\lambda,(m)}_{\mathbf{k}}(\eta) h^{\lambda^{\prime},(n)}_{\mathbf{k}'}(\eta)\rangle$ only has non-zero results when $n = m = 2$ and $n = m = 3$, namely
\begin{eqnarray}\label{eq:12}
 \langle h^{\lambda,(2)}_{\mathbf{k}}(\eta) &&h^{\lambda',(2)}_{\mathbf{k}'}(\eta) \rangle \sim   \langle \zeta_{\mathbf{k}-\mathbf{p}}\zeta_{\mathbf{p}} \zeta_{\mathbf{k}'-\mathbf{p}'}\zeta_{\mathbf{p}'}  \rangle \nonumber\\
  \sim&& \mathcal{P}_{\zeta}\mathcal{P}_{\zeta}  \sim  A_{\zeta}^2 \ , \\
  \langle h^{\lambda,(3)}_{\mathbf{k}}(\eta)&& h^{\lambda',(3)}_{\mathbf{k}'}(\eta) \rangle \sim \langle \zeta_{\mathbf{k}-\mathbf{p}}\zeta_{\mathbf{p}-\mathbf{q}}\zeta_{\mathbf{q}} \zeta_{\mathbf{k}'-\mathbf{p}'}\zeta_{\mathbf{p}'-\mathbf{q}'}\zeta_{\mathbf{q}'} \rangle  \nonumber\\
  \sim&& \mathcal{P}_{\zeta}\mathcal{P}_{\zeta}\mathcal{P}_{\zeta}  \sim  A_{\zeta}^3 \ ,
\end{eqnarray}
where $\mathcal{P}_{\zeta}$ is the power spectrum of primordial curvature perturbation which is defined as $\left\langle\zeta_{\mathbf{k}} \zeta_{\mathbf{k}^{\prime}}\right\rangle=\frac{2 \pi^2}{k^3} \delta\left(\mathbf{k}+\mathbf{k}^{\prime}\right) \mathcal{P}_\zeta(k)$. $A_{\zeta}$ is the amplitude of the primordial power spectrum.

\begin{figure}[htbp]
    \includegraphics[width=1\columnwidth]{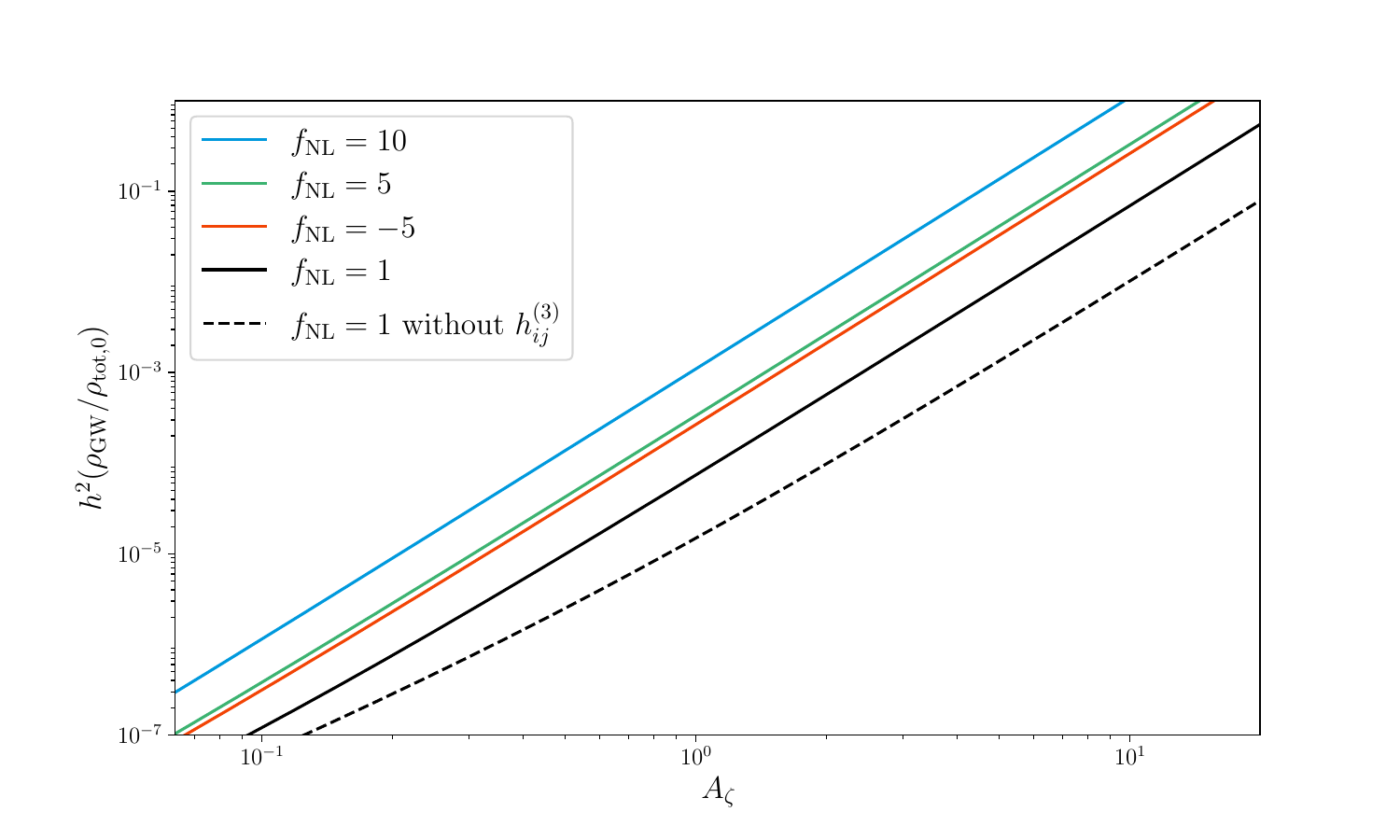}
\caption{\label{rho.pdf} Energy density of SIGWs ($\rho_{\mathrm{GW}} /\rho_{\mathrm{tot},0}$) as a function of $A_\zeta$. Different curves represent varying values of $f_{\rm{NL}}$. The black dashed line denotes the energy density  when only second-order SIGWs $h^{(2)}_{ij}$ are considered. We have set $\sigma=0.5$.}
\end{figure}

In this paper, we consider the local type non-Gaussianity which can be expressed as a local perturbative expansion around the Gaussian primordial curvature perturbation. The primordial curvature perturbation in momentum space can be rewritten as \cite{Cai:2018dig,Adshead:2021hnm,Domenech:2021ztg,Chang:2023aba}
\begin{eqnarray}\label{eq:ngk}
	\zeta^{\mathrm{ng}}_{\mathbf{k}}=\zeta_{\mathbf{k}}+\frac{3}{5}f_{\mathrm{NL}}\int\frac{d^3\mathbf{n}}{(2\pi)^{3/2}} \zeta_{\mathbf{k}-\mathbf{n}}\zeta_{\mathbf{n}}  \ .
\end{eqnarray}
Here, we consider the lowest-order contribution of the non-Gaussian parameter $f_{\mathrm{NL}}$.
In this case, the contribution of the two point correlation function $\langle h^{\lambda,(m)}_{\mathbf{k}}(\eta) h^{\lambda^{\prime},(n)}_{\mathbf{k}'}(\eta)\rangle$ to the total power spectrum $\mathcal{P}_h(\eta,k)$ of SIGWs can be divided into four parts
\begin{eqnarray}\label{eq:1234}
  & &\langle h^{\lambda,(2)}_{\mathbf{k}}(\eta) h^{\lambda',(2)}_{\mathbf{k}'}(\eta) \rangle\sim A_{\zeta}^2 \ , \ \langle h^{\lambda,(3)}_{\mathbf{k}}(\eta) h^{\lambda',(3)}_{\mathbf{k}'}(\eta) \rangle\sim A_{\zeta}^3 \ , \nonumber\\
  & &\langle h^{\lambda,(3)}_{\mathbf{k}}(\eta) h^{\lambda',(2)}_{\mathbf{k}'}(\eta) \rangle\sim f_{\mathrm{NL}}A_{\zeta}^3 \ ,  \nonumber\\
  & & \langle h^{\lambda,(2)}_{\mathbf{k}}(\eta) h^{\lambda',(2)}_{\mathbf{k}'}(\eta) \rangle\sim (f_{\mathrm{NL}})^2A_{\zeta}^3 \ .
\end{eqnarray}
By substituting Eq.~(\ref{eq:H0})--Eq.~(\ref{eq:H4}) and Eq.~(\ref{eq:2h}) into Eq.~(\ref{eq:Ph}), we can derive the expression for the total power spectrum $\mathcal{P}_h(\eta,k)$ of SIGWs. Subsequently, using Eq.~(\ref{eq:Ome}), we can calculate the total energy density spectrum of SIGWs. Taking into account the thermal history of the universe, we obtain the current total energy density spectrum of 
 SIGWs $\Omega_{\mathrm{GW},0}$ \cite{Wang:2019kaf}
\begin{equation}\label{eq:tot_spectrum}
	\begin{aligned}
	\Omega_{\mathrm{GW},0}(k)=\Omega_{\mathrm{rad}, 0}\left(\frac{g_{*, \rho, \mathrm{e}}}{g_{*, \rho, 0}}\right)\left(\frac{g_{*, s, 0}}{g_{*, s, \mathrm{e}}}\right)^{4 / 3} \overline{\Omega}_{\mathrm{GW}}(\eta, k) \ ,
	\end{aligned}
\end{equation}
where $\Omega_{\mathrm{rad}, 0}\approx 4.2\times 10^{-5}/h^2$ is the physical energy density fraction of radiations in the present universe. The effect numbers of relativistic species $g_{*,\rho}$ and $g_{*,s}$ can be found in Ref.~\cite{Saikawa:2018rcs}. And $h =0.6766$ is the dimensionless Hubble constant.
By integrating the energy density spectrum of SIGWs, we obtain the energy density of SIGWs 
\begin{eqnarray}\label{eq:rho}
 \rho_{\mathrm{GW}} /\rho_{\mathrm{tot},0}= \int_0^{\infty} \Omega_{\mathrm{GW},0}(k)d\left(\ln k\right) \ .
\end{eqnarray}
Given the power spectrum of primordial curvature perturbations, we can calculate the energy density spectrum of SIGWs and the corresponding energy density. Here, we consider the log-normal primordial power spectrum
\begin{equation}
    \mathcal{P}_{\zeta}(k) = \frac{A_\zeta}{\sqrt{2\pi\sigma^2}}\exp\left(-\frac{\ln^{2}(k/k_*)}{2\sigma^2}\right) \ ,
\end{equation}
where $A_{\zeta}$ is the amplitude of primordial power spectrum and $k_*=2\pi f_*$ is the wavenumber at which the primordial power spectrum has a log-normal peak.  We calculate the total energy density $\rho_{\mathrm{GW}}/\rho_{\mathrm{tot},0}$ of SIGWs for different $A_{\zeta}$ and $f_{\mathrm{NL}}$. As shown in Fig.~\ref{rho.pdf},  the green and orange lines represent the total energy density of SIGWs for non-Gaussian parameters $f_{\mathrm{NL}}=5$ and $f_{\mathrm{NL}}=-5$, respectively. It shows that the energy density for $f_{\mathrm{NL}}=-5$ is significantly lower than that for $f_{\mathrm{NL}}=5$. The presence of the cross two-point correlation function $\langle h^{\lambda,(3)}_{\mathbf{k}} h^{\lambda',(2)}_{\mathbf{k}'}\rangle\sim f_{\mathrm{NL}}A_{\zeta}^3$ causes a significant decrease in the total energy density spectrum of SIGWs when the non-Gaussian parameter $f_{\mathrm{NL}}$ is negative. When $f_{\mathrm{NL}}$ is sufficiently large, the contribution of the two-point correlation function $\langle h^{\lambda,(2)}_{\mathbf{k}} h^{\lambda',(2)}_{\mathbf{k}'}\rangle\sim \left(f_{\mathrm{NL}}\right)^2A_{\zeta}^3$ becomes dominant, and the suppressing effect of the cross two-point correlation function $\langle h^{\lambda,(3)}_{\mathbf{k}} h^{\lambda',(2)}_{\mathbf{k}'}\rangle\sim f_{\mathrm{NL}}A_{\zeta}^3$ on the total energy density becomes negligible  \cite{Chang:2023aba}. Additionally, the black dashed line in fig.~\ref{rho.pdf} represents the energy density spectrum with second-order SIGWs considered only. The difference between the black solid line and the black dashed line reveals that third-order SIGWs substantially increase the total energy density of SIGWs.

\section{Constraints on \acp{PBH}} \label{sec:4.0}
When the primordial curvature perturbations on small scales are sufficiently large, \acp{PBH} will inevitably form. The \ac{PBH} mass fraction at formation time can be calculated as \cite{Young:2013oia,Byrnes:2012yx}
\begin{eqnarray}\label{eq:beta}
    \beta(M) \approx \frac{1}{2}\left\{\begin{array}{cl}
\operatorname{erfc}\left(\frac{\zeta_{\mathrm{G}}^{+}\left(\zeta_{\mathrm{c}}\right)}{\sqrt{2\left\langle\zeta_{\mathrm{G}}^2\right\rangle}}\right)+\operatorname{erfc}\left(-\frac{\zeta_{\mathrm{G}}^{-}\left(\zeta_{\mathrm{c}}\right)}{\sqrt{2\left\langle\zeta_{\mathrm{G}}^2\right\rangle}}\right) ; \\
 f_{\mathrm{NL}}>0 \ , \\
\operatorname{erf}\left(\frac{\zeta_{\mathrm{G}}^{+}\left(\zeta_{\mathrm{c}}\right)}{\sqrt{2\left\langle\zeta_{\mathrm{G}}^2\right\rangle}}\right)-\operatorname{erf}\left(\frac{\zeta_{\mathrm{G}}^{-}\left(\zeta_{\mathrm{c}}\right)}{\sqrt{2\left\langle\zeta_{\mathrm{G}}^2\right\rangle}}\right) ; \\
 f_{\mathrm{NL}}<0\ ,
\end{array}\right.
\end{eqnarray}
where
\begin{eqnarray}
    \zeta_{G}^{\pm}(\zeta)\approx \frac{5}{6f_{\mathrm{NL}}}\left( -1\pm \sqrt{1+\frac{12}{5}f_{\mathrm{NL}}\zeta} \right) \ .
\end{eqnarray}
And $\zeta_c$  is the threshold value \cite{Musco:2008hv,Musco:2004ak,Harada:2013epa}.  The abundance of \acp{PBH} is given by \cite{Sasaki:2018dmp}
\begin{eqnarray}\label{eq:fbeta}
   f_{\mathrm{pbh}} \simeq 2.5 \times 10^8 \beta\left(\frac{g_*^{\text {form }}}{10.75}\right)^{-\frac{1}{4}}\left(\frac{m_{\mathrm{pbh}}}{M_{\odot}}\right)^{-\frac{1}{2}} \ .
\end{eqnarray}
Given the primordial power spectrum, we can use Eq.~(\ref{eq:beta}) and Eq.~(\ref{eq:fbeta})  to calculate the abundance of \acp{PBH}. In the subsequent discussion, we will analyze the constraints that current $N_{\text{eff}}$ observational data impose on both the primordial power spectrum and the abundance of \acp{PBH}.

\begin{figure}[htbp]
    \includegraphics[width=1.1\columnwidth]{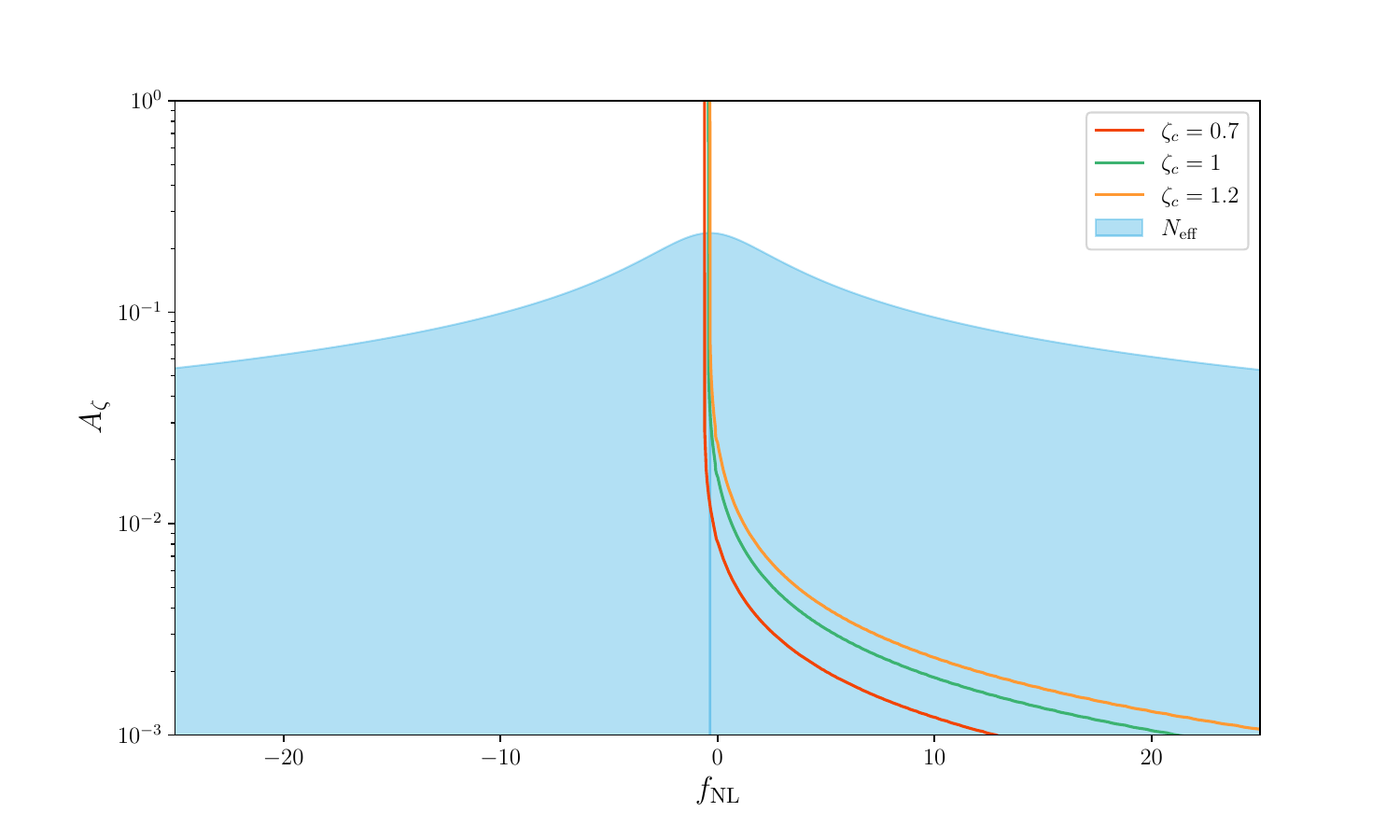}
\caption{\label{Neff.pdf} Constraints on $A_\zeta$ under different values of $f_{\rm{NL}}$. The shaded blue region represents the allowed area constrained by current observations of $N_{\rm{eff}}$. The three curves indicate the upper limits of $A_\zeta$ for different values of $f_{\rm{NL}}$ under the condition $f_{\rm{pbh}} = 1$ and $m_{\rm{pbh}}=10^{-12}M_{\odot}$, 
 where $M_{\odot}$ denotes the solar mass. We have chosen the threshold value $\zeta_c$ to be $0.7$, $1$, and $1.2$, respectively.}
\end{figure}

After the decoupling of neutrinos, the cosmic radiation energy density $\rho_{\mathrm{rad}}$ includes \ac{CMB} photons $\gamma$, neutrinos $\nu$, and the density of SIGWs
\begin{eqnarray}
    \rho_{\mathrm{rad}}=\rho_{\gamma}+\rho_{\nu}+\rho_{\mathrm{GW}} \ .
\end{eqnarray}
The relationship between the three energy densities, \ac{CMB} temperature, and the relativistic degrees of freedom $N_{\text{eff}}$ can be expressed as
\begin{eqnarray}
    \rho_{\gamma}&=&\frac{\pi^2}{15}T_{\gamma}^4  \ , \label{eq:Tg}\\
    \rho_{\nu}+\rho_{\mathrm{GW}}&=&\frac{7\pi^2}{120}N_{\text{eff}}T_{\nu}^4 \ , \label{eq:Tv}
\end{eqnarray}
where
$T_{\gamma}=2.728(1+z)$ K and $T_{\nu}=(4/11)^{1/3}T_{\gamma}$ K are temperatures of \ac{CMB} and neutrino respectively. The energy density of SIGWs can affect $N_{\text{eff}}$, and the upper limit of this effect is constrained by current cosmological observations. In this case, the total energy density spectrum of SIGWs satisfies \cite{Ben-Dayan:2019gll,Clarke:2020bil,Cang:2022jyc,Wang:2023sij,Wright:2024awr}
\begin{eqnarray}\label{eq:rhup}
  \int_{f_{\mathrm{min}}}^{\infty} h^2\Omega_{\mathrm{GW},0}(k) d\left(\ln k\right) < 1.3 \times 10^{-6}\frac{\Delta N_{\text{eff}}}{0.234} \    .
\end{eqnarray}
 where $\Delta N_{\text{eff}}= N_{\text{eff}}-3.046$. We utilize the $N_{\text{eff}}$ limits provided by \cite{Planck:2018vyg}, which report $N_{\text{eff}}=3.04 \pm 0.22$ at a $95\%$ confidence level for the \texttt{ Planck} + BAO + BBN data. By applying Gaussian statistics, this translates to a $95\%$ confidence level upper limit of $\Delta N_{\text{eff}}<0.175$. Using Eq.~(\ref{eq:rhup}), we can provide the current cosmological constraints on the primordial power spectrum based on observations of $N_{\text{eff}}$. As shown in Fig.~\ref{Neff.pdf}, we present the current constraints on the small-scale primordial power spectrum based on the \texttt{ Planck} + BAO + BBN data \cite{Planck:2018vyg}.
In Fig.~\ref{Neff.pdf}, the three curves represent the relationship between parameter $A_{\zeta}$ and $f_{\mathrm{NL}}$ when the primordial black hole abundance $f_{\mathrm{pbh}}$ is set to $1$.  It shows that for different non-Gaussian parameters $f_{\mathrm{NL}}$ , current $N_{\rm{eff}}$ observations can effectively constrain the amplitude of the small-scale primordial power spectrum $A_{\zeta}$. For the abundance of primordial black holes, the current $N_{\rm{eff}}$ observations can only effectively limit the abundance of primordial black holes with a mass around $m_{\rm{pbh}}=10^{-12}M_{\odot}$ when the non-Gaussian parameter $f_{\mathrm{pbh}}$ is less than zero and within a narrow range near zero. For $f_{\mathrm{NL}} > 0$ values, the $f_{\mathrm{pbh}}=1$ curves fall entirely within the allowable range of current $N_{\rm{eff}}$ observations. 

\begin{figure}[htbp]
    \includegraphics[width=0.95\columnwidth]{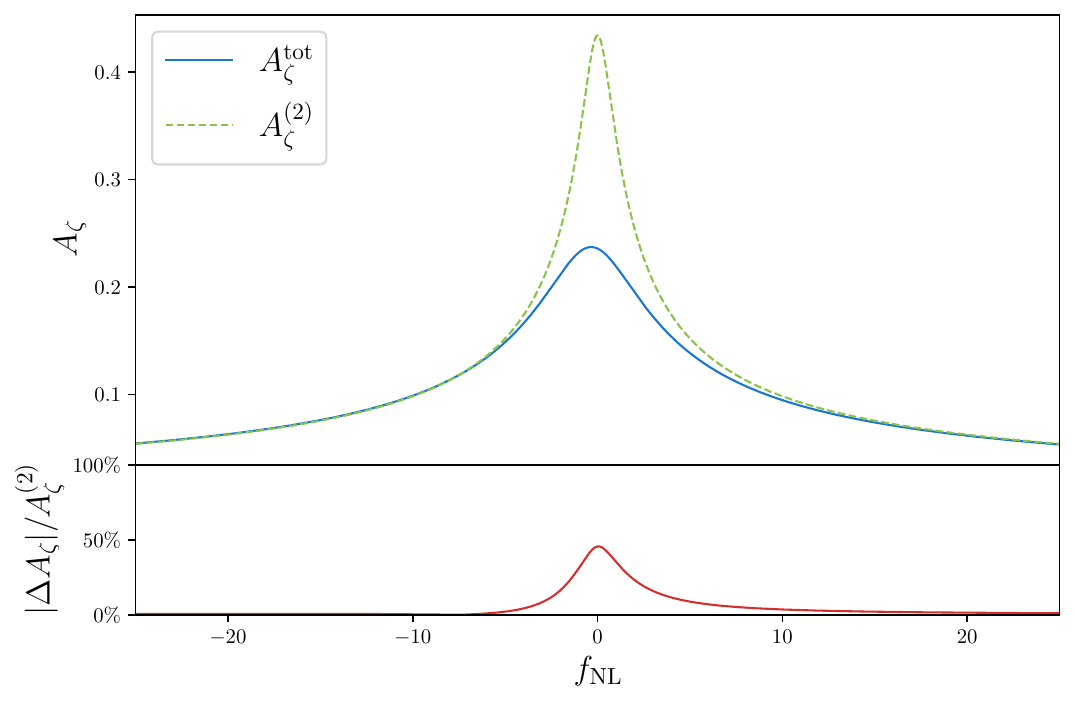}
\caption{\label{DeltaA.pdf} Constraints on $A_\zeta$ under different values of $f_{\rm{NL}}$.  Symbol $A_{\zeta}^{(2)}$ denotes the upper limit for $A_{\zeta}$ determined from $N_{\rm{eff}}$ observations, taking into account only the second-order SIGWs. Symbol $A_{\zeta}^{\mathrm{tot}}$ denotes the upper limit for $A_{\zeta}$ determined from $N_{\rm{eff}}$ observations, considering all two-point correlation functions in Eq.~(\ref{eq:1234}). $\Delta A_{\zeta}=A_{\zeta}^{(2)}-A_{\zeta}^{\mathrm{tot}}$ denotes the difference between the two upper bounds of $A_{\zeta}$. }
\end{figure}

To more clearly demonstrate the effect of third-order SIGWs on current $N_{\rm{eff}}$ observations, we provide a comparative illustration of the upper limits of $A_{\zeta}$ based on $N_{\rm{eff}}$ observational data, with and without the influence of third-order SIGWs. As shown in Fig.~\ref{DeltaA.pdf}, the solid blue line represents the upper limit of $A_{\zeta}$ when third-order SIGWs are considered. The dashed green line represents the upper limit of $A_{\zeta}$ when the two-point correlation functions $\langle h^{\lambda,(3)}_{\mathbf{k}} h^{\lambda',(2)}_{\mathbf{k}'} \rangle$ and $\langle h^{\lambda,(3)}_{\mathbf{k}} h^{\lambda',(3)}_{\mathbf{k}'} \rangle$ are neglected. Our results indicate that for small absolute values of $f_{\mathrm{NL}}$, third-order SIGWs greatly reduce the upper limit of $A_{\zeta}$. For larger absolute values of $f_{\mathrm{NL}}$, the non-Gaussian contribution from second-order SIGWs becomes dominant, and the influence of third-order SIGWs on the total energy density is comparatively minor. As the absolute value of $f_{\mathrm{NL}}$ increases, the blue solid line and the green dashed line in Fig.~\ref{DeltaA.pdf} gradually coincide. Furthermore, unlike the scenario considering only second-order SIGWs, due to the presence of the cross-correlation function $\langle h^{\lambda,(3)}_{\mathbf{k}} h^{\lambda',(2)}_{\mathbf{k}'} \rangle \sim f_{\mathrm{NL}}A^{3}_{\zeta}$, the blue solid line in Fig.~\ref{DeltaA.pdf} is not symmetric about $f_{\mathrm{NL}}=0$.

\section{Conclusion and discussion} \label{sec:5.0}
We investigated the impact of SIGWs as an additional radiation component on cosmological evolution. By utilizing the parameter $N_{\text{eff}}$ from current large-scale cosmological observations, we can place stringent limits on the energy density of these waves. This allows us to constrain the parameter space of the small-scale primordial power spectrum and \acp{PBH}. In this paper, we calculated the total energy density of SIGWs up to $A_{\zeta}^3$ order, taking into account a non-Gaussian primordial power spectrum. Our results indicate that the parameter space of the primordial power spectrum can be significantly constrained. Additionally,  the current $N_{\text{eff}}$ observations can effectively constrain the primordial black hole abundance only if the non-Gaussian parameter $f_{\mathrm{NL}}$ is less than zero and within a narrow range near zero. For $f_{\mathrm{NL}}>0$, the current $N_{\text{eff}}$ observations cannot effectively constrain the abundance of \acp{PBH}.

 To more effectively limit the parameter space of the small-scale primordial power spectrum and \acp{PBH}, future more precise $N_{\text{eff}}$ observational data can be considered. A decrease in $\Delta N_{\text{eff}}$ will inevitably result in a reduced parameter space for the primordial power spectrum and \acp{PBH}. Furthermore, the influence of additional radiation components, apart from SIGWs, can also be taken into account \cite{Ghoshal:2023phi,RoyChoudhury:2022rva,Gerbino:2022nvz,Luo:2020sho,EscuderoAbenza:2020cmq,Bennett:2019ewm}. In this scenario, SIGWs and other extra radiation components will collectively impact the $N_{\text{eff}}$ parameter. Current $N_{\text{eff}}$ observational data can be utilized to constrain the parameter space of multiple extra radiation components simultaneously. These topics may be explored in future research.

In this paper, we consider the contributions of the lowest-order non-Gaussian primordial perturbations to the total energy density spectrum of SIGWs. 
When fully accounting for non-Gaussian primordial perturbations, an additional term should be included in Eq.~(\ref{eq:1234}) \cite{Perna:2024ehx,Chang:2023aba}: $\langle h^{\lambda,(3)}_{\mathbf{k}} h^{\lambda',(2)}_{\mathbf{k}'} \rangle\sim f_{\mathrm{NL}}A_{\zeta}^3+(f_{\mathrm{NL}})^3A_{\zeta}^4+(f_{\mathrm{NL}})^5A_{\zeta}^5$ and $\langle h^{\lambda,(2)}_{\mathbf{k}} h^{\lambda',(2)}_{\mathbf{k}'} \rangle\sim (f_{\mathrm{NL}})^2A_{\zeta}^3+(f_{\mathrm{NL}})^4A_{\zeta}^4$.
Furthermore, here we only consider the contributions of SIGWs up to the third order. When the amplitude $A_{\zeta}$ of the small-scale primordial power spectrum is very large, the contributions of higher-order SIGWs cannot be ignored \cite{Zhou:2024ncc}. These contributions of higher-order non-Gaussianity and higher-order SIGWs will significantly enhance the energy density spectrum of SIGWs, thereby making the constraints of $N_{\text{eff}}$ on the primordial power spectrum and the parameter space of \acp{PBH} more stringent.  In this study, we investigated the contributions to the energy density spectrum of SIGWs up to the $A^{3}_{\zeta}$ order. The resulting constraints can still be used as an upper limit for the abundance of \acp{PBH}. Future research may further refine the study of higher-order contributions.

\renewcommand\linenumberfont{\normalfont\small\color{white}}
\begin{acknowledgments}
The work is supported in part by the National Natural Science Foundation of China (NSFC) grants  No.12475075, No.11935009, No.12375052, No.12075249, No.11690022, No.12275276, and No.12447127.
\end{acknowledgments}

\bibliography{biblio}{}

\begin{thebibliography}{}
\expandafter\ifx\csname natexlab\endcsname\relax\def\natexlab#1{#1}\fi
\providecommand{\url}[1]{\href{#1}{#1}}
\providecommand{\dodoi}[1]{doi:~\href{http://doi.org/#1}{\nolinkurl{#1}}}
\providecommand{\doeprint}[1]{\href{http://ascl.net/#1}{\nolinkurl{http://ascl.net/#1}}}
\providecommand{\doarXiv}[1]{\href{https://arxiv.org/abs/#1}{\nolinkurl{https://arxiv.org/abs/#1}}}

\bibitem[{Abdalla {et~al.}(2022)}]{Abdalla:2022yfr}
Abdalla, E., {et~al.} 2022, JHEAp, 34, 49, \dodoi{10.1016/j.jheap.2022.04.002}

\bibitem[{Adshead {et~al.}(2021)Adshead, Lozanov, \& Weiner}]{Adshead:2021hnm}
Adshead, P., Lozanov, K.~D., \& Weiner, Z.~J. 2021, JCAP, 10, 080, \dodoi{10.1088/1475-7516/2021/10/080}

\bibitem[{Agazie {et~al.}(2023)}]{NANOGrav:2023gor}
Agazie, G., {et~al.} 2023, Astrophys. J. Lett., 951, L8, \dodoi{10.3847/2041-8213/acdac6}

\bibitem[{Aghanim {et~al.}(2020)}]{Planck:2018vyg}
Aghanim, N., {et~al.} 2020, Astron. Astrophys., 641, A6, \dodoi{10.1051/0004-6361/201833910}

\bibitem[{Ananda {et~al.}(2007)Ananda, Clarkson, \& Wands}]{Ananda:2006af}
Ananda, K.~N., Clarkson, C., \& Wands, D. 2007, Phys. Rev. D, 75, 123518, \dodoi{10.1103/PhysRevD.75.123518}

\bibitem[{Antoniadis {et~al.}(2023)}]{EPTA:2023fyk}
Antoniadis, J., {et~al.} 2023, Astron. Astrophys., 678, A50, \dodoi{10.1051/0004-6361/202346844}

\bibitem[{Barack {et~al.}(2019)}]{Barack:2018yly}
Barack, L., {et~al.} 2019, Class. Quant. Grav., 36, 143001, \dodoi{10.1088/1361-6382/ab0587}

\bibitem[{Baumann {et~al.}(2007)Baumann, Steinhardt, Takahashi, \& Ichiki}]{Baumann:2007zm}
Baumann, D., Steinhardt, P.~J., Takahashi, K., \& Ichiki, K. 2007, Phys. Rev. D, 76, 084019, \dodoi{10.1103/PhysRevD.76.084019}

\bibitem[{Ben-Dayan {et~al.}(2019)Ben-Dayan, Keating, Leon, \& Wolfson}]{Ben-Dayan:2019gll}
Ben-Dayan, I., Keating, B., Leon, D., \& Wolfson, I. 2019, JCAP, 06, 007, \dodoi{10.1088/1475-7516/2019/06/007}

\bibitem[{Bennett {et~al.}(2020)Bennett, Buldgen, Drewes, \& Wong}]{Bennett:2019ewm}
Bennett, J.~J., Buldgen, G., Drewes, M., \& Wong, Y. Y.~Y. 2020, JCAP, 03, 003, \dodoi{10.1088/1475-7516/2020/03/003}

\bibitem[{Bird {et~al.}(2016)Bird, Cholis, Mu\~noz, Ali-Ha\"\i{}moud, Kamionkowski, Kovetz, Raccanelli, \& Riess}]{Bird:2016dcv}
Bird, S., Cholis, I., Mu\~noz, J.~B., {et~al.} 2016, Phys. Rev. Lett., 116, 201301, \dodoi{10.1103/PhysRevLett.116.201301}

\bibitem[{Bringmann {et~al.}(2012)Bringmann, Scott, \& Akrami}]{Bringmann:2011ut}
Bringmann, T., Scott, P., \& Akrami, Y. 2012, Phys. Rev. D, 85, 125027, \dodoi{10.1103/PhysRevD.85.125027}

\bibitem[{Bugaev \& Klimai(2010)}]{Bugaev:2009zh}
Bugaev, E., \& Klimai, P. 2010, Phys. Rev. D, 81, 023517, \dodoi{10.1103/PhysRevD.81.023517}

\bibitem[{Bugaev \& Klimai(2011)}]{Bugaev:2010bb}
---. 2011, Phys. Rev. D, 83, 083521, \dodoi{10.1103/PhysRevD.83.083521}

\bibitem[{Byrnes {et~al.}(2019)Byrnes, Cole, \& Patil}]{Byrnes:2018txb}
Byrnes, C.~T., Cole, P.~S., \& Patil, S.~P. 2019, JCAP, 06, 028, \dodoi{10.1088/1475-7516/2019/06/028}

\bibitem[{Byrnes {et~al.}(2012)Byrnes, Copeland, \& Green}]{Byrnes:2012yx}
Byrnes, C.~T., Copeland, E.~J., \& Green, A.~M. 2012, Phys. Rev. D, 86, 043512, \dodoi{10.1103/PhysRevD.86.043512}

\bibitem[{Cai {et~al.}(2019)Cai, Pi, \& Sasaki}]{Cai:2018dig}
Cai, R.-g., Pi, S., \& Sasaki, M. 2019, Phys. Rev. Lett., 122, 201101, \dodoi{10.1103/PhysRevLett.122.201101}

\bibitem[{Cang {et~al.}(2023)Cang, Ma, \& Gao}]{Cang:2022jyc}
Cang, J., Ma, Y.-Z., \& Gao, Y. 2023, Astrophys. J., 949, 64, \dodoi{10.3847/1538-4357/acc949}

\bibitem[{Carr {et~al.}(2021)Carr, Kohri, Sendouda, \& Yokoyama}]{Carr:2020gox}
Carr, B., Kohri, K., Sendouda, Y., \& Yokoyama, J. 2021, Rept. Prog. Phys., 84, 116902, \dodoi{10.1088/1361-6633/ac1e31}

\bibitem[{Chang {et~al.}(2024{\natexlab{a}})Chang, Kuang, Wu, \& Zhou}]{Chang:2023vjk}
Chang, Z., Kuang, Y.-T., Wu, D., \& Zhou, J.-Z. 2024{\natexlab{a}}, JCAP, 2024, 044, \dodoi{10.1088/1475-7516/2024/04/044}

\bibitem[{Chang {et~al.}(2024{\natexlab{b}})Chang, Kuang, Wu, Zhou, \& Zhu}]{Chang:2023aba}
Chang, Z., Kuang, Y.-T., Wu, D., Zhou, J.-Z., \& Zhu, Q.-H. 2024{\natexlab{b}}, Phys. Rev. D, 109, L041303, \dodoi{10.1103/PhysRevD.109.L041303}

\bibitem[{Chang {et~al.}(2023)Chang, Kuang, Zhang, \& Zhou}]{Chang:2022nzu}
Chang, Z., Kuang, Y.-T., Zhang, X., \& Zhou, J.-Z. 2023, Chin. Phys. C, 47, 055104, \dodoi{10.1088/1674-1137/acc649}

\bibitem[{Chen {et~al.}(2024)Chen, Ghoshal, Tasinato, \& Tomberg}]{Chen:2024pge}
Chen, C., Ghoshal, A., Tasinato, G., \& Tomberg, E. 2024.
\newblock \doarXiv{2409.12950}

\bibitem[{Chen {et~al.}(2023)Chen, Ota, Zhu, \& Zhu}]{Chen:2022dah}
Chen, C., Ota, A., Zhu, H.-Y., \& Zhu, Y. 2023, Phys. Rev. D, 107, 083518, \dodoi{10.1103/PhysRevD.107.083518}

\bibitem[{Choudhury {et~al.}(2024)Choudhury, Ganguly, Panda, SenGupta, \& Tiwari}]{Choudhury:2024dzw}
Choudhury, S., Ganguly, S., Panda, S., SenGupta, S., \& Tiwari, P. 2024, JCAP, 09, 013, \dodoi{10.1088/1475-7516/2024/09/013}

\bibitem[{Clarke {et~al.}(2020)Clarke, Copeland, \& Moss}]{Clarke:2020bil}
Clarke, T.~J., Copeland, E.~J., \& Moss, A. 2020, JCAP, 10, 002, \dodoi{10.1088/1475-7516/2020/10/002}

\bibitem[{Dom\`enech(2021)}]{Domenech:2021ztg}
Dom\`enech, G. 2021, Universe, 7, 398, \dodoi{10.3390/universe7110398}

\bibitem[{Dom\`enech {et~al.}(2024)Dom\`enech, Pi, Wang, \& Wang}]{Domenech:2024rks}
Dom\`enech, G., Pi, S., Wang, A., \& Wang, J. 2024, JCAP, 08, 054, \dodoi{10.1088/1475-7516/2024/08/054}

\bibitem[{Dom\`enech \& Tr\"ankle(2024)}]{Domenech:2024wao}
Dom\`enech, G., \& Tr\"ankle, J. 2024.
\newblock \doarXiv{2409.12125}

\bibitem[{Ellis {et~al.}(2024)Ellis, Fairbairn, Franciolini, H\"utsi, Iovino, Lewicki, Raidal, Urrutia, Vaskonen, \& Veerm\"ae}]{Ellis:2023oxs}
Ellis, J., Fairbairn, M., Franciolini, G., {et~al.} 2024, Phys. Rev. D, 109, 023522, \dodoi{10.1103/PhysRevD.109.023522}

\bibitem[{Escudero~Abenza(2020)}]{EscuderoAbenza:2020cmq}
Escudero~Abenza, M. 2020, JCAP, 05, 048, \dodoi{10.1088/1475-7516/2020/05/048}

\bibitem[{Franciolini {et~al.}(2023)Franciolini, Iovino, Vaskonen, \& Veermae}]{Franciolini:2023pbf}
Franciolini, G., Iovino, Junior., A., Vaskonen, V., \& Veermae, H. 2023, Phys. Rev. Lett., 131, 201401, \dodoi{10.1103/PhysRevLett.131.201401}

\bibitem[{Garc\'\i{}a-Bellido(2017)}]{Garcia-Bellido:2017fdg}
Garc\'\i{}a-Bellido, J. 2017, J. Phys. Conf. Ser., 840, 012032, \dodoi{10.1088/1742-6596/840/1/012032}

\bibitem[{Gerbino {et~al.}(2023)}]{Gerbino:2022nvz}
Gerbino, M., {et~al.} 2023, Phys. Dark Univ., 42, 101333, \dodoi{10.1016/j.dark.2023.101333}

\bibitem[{Ghoshal {et~al.}(2023)Ghoshal, Lalak, \& Porey}]{Ghoshal:2023phi}
Ghoshal, A., Lalak, Z., \& Porey, S. 2023, Phys. Rev. D, 108, 063030, \dodoi{10.1103/PhysRevD.108.063030}

\bibitem[{Harada {et~al.}(2013)Harada, Yoo, \& Kohri}]{Harada:2013epa}
Harada, T., Yoo, C.-M., \& Kohri, K. 2013, Phys. Rev. D, 88, 084051, \dodoi{10.1103/PhysRevD.88.084051}

\bibitem[{Harigaya {et~al.}(2023)Harigaya, Inomata, \& Terada}]{Harigaya:2023pmw}
Harigaya, K., Inomata, K., \& Terada, T. 2023, Phys. Rev. D, 108, 123538, \dodoi{10.1103/PhysRevD.108.123538}

\bibitem[{Hosseini~Mansoori {et~al.}(2023)Hosseini~Mansoori, Felegray, Talebian, \& Sami}]{HosseiniMansoori:2023mqh}
Hosseini~Mansoori, S.~A., Felegray, F., Talebian, A., \& Sami, M. 2023, JCAP, 08, 067, \dodoi{10.1088/1475-7516/2023/08/067}

\bibitem[{Jiang {et~al.}(2024)Jiang, Cai, Ye, \& Piao}]{Jiang:2023gfe}
Jiang, J.-Q., Cai, Y., Ye, G., \& Piao, Y.-S. 2024, JCAP, 05, 004, \dodoi{10.1088/1475-7516/2024/05/004}

\bibitem[{Kohri \& Terada(2018)}]{Kohri:2018awv}
Kohri, K., \& Terada, T. 2018, Phys. Rev. D, 97, 123532, \dodoi{10.1103/PhysRevD.97.123532}

\bibitem[{Luo {et~al.}(2020)Luo, Rodejohann, \& Xu}]{Luo:2020sho}
Luo, X., Rodejohann, W., \& Xu, X.-J. 2020, JCAP, 06, 058, \dodoi{10.1088/1475-7516/2020/06/058}

\bibitem[{Matarrese {et~al.}(1998)Matarrese, Mollerach, \& Bruni}]{Matarrese:1997ay}
Matarrese, S., Mollerach, S., \& Bruni, M. 1998, Phys. Rev. D, 58, 043504, \dodoi{10.1103/PhysRevD.58.043504}

\bibitem[{Mollerach {et~al.}(2004)Mollerach, Harari, \& Matarrese}]{Mollerach:2003nq}
Mollerach, S., Harari, D., \& Matarrese, S. 2004, Phys. Rev. D, 69, 063002, \dodoi{10.1103/PhysRevD.69.063002}

\bibitem[{Musco {et~al.}(2009)Musco, Miller, \& Polnarev}]{Musco:2008hv}
Musco, I., Miller, J.~C., \& Polnarev, A.~G. 2009, Class. Quant. Grav., 26, 235001, \dodoi{10.1088/0264-9381/26/23/235001}

\bibitem[{Musco {et~al.}(2005)Musco, Miller, \& Rezzolla}]{Musco:2004ak}
Musco, I., Miller, J.~C., \& Rezzolla, L. 2005, Class. Quant. Grav., 22, 1405, \dodoi{10.1088/0264-9381/22/7/013}

\bibitem[{Papanikolaou(2023)}]{Papanikolaou:2023oxq}
Papanikolaou, T. 2023, PoS, CORFU2022, 265, \dodoi{10.22323/1.436.0265}

\bibitem[{Papanikolaou {et~al.}(2024)Papanikolaou, Banerjee, Cai, Capozziello, \& Saridakis}]{Papanikolaou:2024fzf}
Papanikolaou, T., Banerjee, S., Cai, Y.-F., Capozziello, S., \& Saridakis, E.~N. 2024, JCAP, 06, 066, \dodoi{10.1088/1475-7516/2024/06/066}

\bibitem[{Papanikolaou {et~al.}(2021)Papanikolaou, Vennin, \& Langlois}]{Papanikolaou:2020qtd}
Papanikolaou, T., Vennin, V., \& Langlois, D. 2021, JCAP, 03, 053, \dodoi{10.1088/1475-7516/2021/03/053}

\bibitem[{Perna {et~al.}(2024)Perna, Testini, Ricciardone, \& Matarrese}]{Perna:2024ehx}
Perna, G., Testini, C., Ricciardone, A., \& Matarrese, S. 2024, JCAP, 05, 086, \dodoi{10.1088/1475-7516/2024/05/086}

\bibitem[{Reardon {et~al.}(2023)}]{Reardon:2023gzh}
Reardon, D.~J., {et~al.} 2023, Astrophys. J. Lett., 951, L6, \dodoi{10.3847/2041-8213/acdd02}

\bibitem[{Roy~Choudhury {et~al.}(2022)Roy~Choudhury, Hannestad, \& Tram}]{RoyChoudhury:2022rva}
Roy~Choudhury, S., Hannestad, S., \& Tram, T. 2022, JCAP, 10, 018, \dodoi{10.1088/1475-7516/2022/10/018}

\bibitem[{Saikawa \& Shirai(2018)}]{Saikawa:2018rcs}
Saikawa, K., \& Shirai, S. 2018, JCAP, 05, 035, \dodoi{10.1088/1475-7516/2018/05/035}

\bibitem[{Saito \& Yokoyama(2009)}]{Saito:2008jc}
Saito, R., \& Yokoyama, J. 2009, Phys. Rev. Lett., 102, 161101, \dodoi{10.1103/PhysRevLett.102.161101}

\bibitem[{Sasaki {et~al.}(2018)Sasaki, Suyama, Tanaka, \& Yokoyama}]{Sasaki:2018dmp}
Sasaki, M., Suyama, T., Tanaka, T., \& Yokoyama, S. 2018, Class. Quant. Grav., 35, 063001, \dodoi{10.1088/1361-6382/aaa7b4}

\bibitem[{Sui {et~al.}(2024)Sui, Liu, Yang, \& Cai}]{Sui:2024nip}
Sui, X.-B., Liu, J., Yang, X.-Y., \& Cai, R.-G. 2024.
\newblock \doarXiv{2407.04220}

\bibitem[{Wang {et~al.}(2019)Wang, Terada, \& Kohri}]{Wang:2019kaf}
Wang, S., Terada, T., \& Kohri, K. 2019, Phys. Rev. D, 99, 103531, \dodoi{10.1103/PhysRevD.99.103531}

\bibitem[{Wang {et~al.}(2024)Wang, Zhao, \& Zhu}]{Wang:2023sij}
Wang, S., Zhao, Z.-C., \& Zhu, Q.-H. 2024, Phys. Rev. Res., 6, 013207, \dodoi{10.1103/PhysRevResearch.6.013207}

\bibitem[{Wright {et~al.}(2024)Wright, Giblin, \& Hazboun}]{Wright:2024awr}
Wright, D., Giblin, J.~T., \& Hazboun, J. 2024.
\newblock \doarXiv{2409.15572}

\bibitem[{Xu {et~al.}(2023)}]{Xu:2023wog}
Xu, H., {et~al.} 2023, Res. Astron. Astrophys., 23, 075024, \dodoi{10.1088/1674-4527/acdfa5}

\bibitem[{Young \& Byrnes(2013)}]{Young:2013oia}
Young, S., \& Byrnes, C.~T. 2013, JCAP, 08, 052, \dodoi{10.1088/1475-7516/2013/08/052}

\bibitem[{Yu \& Wang(2024)}]{Yu:2024xmz}
Yu, Y.-H., \& Wang, S. 2024.
\newblock \doarXiv{2405.02960}

\bibitem[{Yuan {et~al.}(2019)Yuan, Chen, \& Huang}]{Yuan:2019udt}
Yuan, C., Chen, Z.-C., \& Huang, Q.-G. 2019, Phys. Rev. D, 100, 081301, \dodoi{10.1103/PhysRevD.100.081301}

\bibitem[{Zhou {et~al.}(2024{\natexlab{a}})Zhou, Kuang, Wu, Chen, L\"u, \& Chang}]{Zhou:2024doz}
Zhou, J.-Z., Kuang, Y.-T., Wu, D., {et~al.} 2024{\natexlab{a}}.
\newblock \doarXiv{2409.07702}

\bibitem[{Zhou {et~al.}(2024{\natexlab{b}})Zhou, Kuang, Wu, L\"u, \& Chang}]{Zhou:2024ncc}
Zhou, J.-Z., Kuang, Y.-T., Wu, D., L\"u, H., \& Chang, Z. 2024{\natexlab{b}}.
\newblock \doarXiv{2408.14052}

\bibitem[{Zhou {et~al.}(2022)Zhou, Zhang, Zhu, \& Chang}]{Zhou:2021vcw}
Zhou, J.-Z., Zhang, X., Zhu, Q.-H., \& Chang, Z. 2022, JCAP, 05, 013, \dodoi{10.1088/1475-7516/2022/05/013}

\end{thebibliography}

\end{document}